\documentclass[3pt,onecolumn]{article}
\usepackage{graphicx}
\usepackage[table]{xcolor}
\definecolor{light-gray}{gray}{0.94}

\newcommand{\ben}{\begin{enumerate}}
\newcommand{\een}{\end{enumerate}}

\newcommand{\be}{\begin{equation}}
\newcommand{\ee}{\end{equation}}
\newcommand{\bea}{\begin{eqnarray}}
\newcommand{\eea}{\end{eqnarray}}

\begin{document}
\begin{flushright}
%Not for distribution, March 2010\\
%Not for further distribution\\
\end{flushright}
\vspace{0.1cm}
\thispagestyle{empty}

\begin{center}
%{\Large\bf }\\[13mm]
{\Large\bf 
Transcription and the Pitch Angle of DNA
}\\[13mm]
%xxxxxxx12345678902234567890323456789042345678905234567890623456789072345678908234567890923456
{\rm Kasper W. Olsen\footnote{DTU Nanotech, Technical University of Denmark, Building 345E, 2800 Kongens Lyngby, Denmark.  E-mail: kasper.olsen@nanotech.dtu.dk} and Jakob Bohr\footnote{DTU Nanotech, Technical University of Denmark, Building 345E, 2800 Kongens Lyngby, Denmark.  E-mail: jabo@nanotech.dtu.dk}}\\[18mm]
\end{center}

\begin{abstract}
The question of the value of the pitch angle of DNA is visited from the perspective of a geometrical analysis of transcription.
It is suggested that for transcription to be possible, the pitch angle of B-DNA must be smaller than the angle of zero-twist.
At the zero-twist angle the double helix is maximally rotated and its strain-twist coupling vanishes.
A numerical estimate of the pitch angle for B-DNA based on differential geometry is compared with numbers obtained from 
existing empirical data. The crystallographic studies shows that the pitch angle is approximately 38$^\circ$, less than the corresponding zero-twist angle of $41.8^\circ$, which is consistent with the suggested principle for transcription.\\[7mm] 
\end{abstract}
%\newpage

\addtocounter{section}{1}

In this paper,  the restrictions inferred on the double-stranded DNA are studied in a geometrical analysis of the mechanics of transcription. 
Transcription of DNA is the biological process where the base-pair code within a gene is read for the purpose of protein expression. The precise mechanisms that control transcription are still to be fully explored, and represent a fundamental theme of research in molecular biology. 
In eucaryotes, one fundamental control point for the regulation of transcription is the structural adaptation of the chromatin fiber necessary to access the gene sequence. One scenario that has been proposed is based on the reversome, explaining how transcription elongation can proceed within condensed chromatin  \cite{becavin2010}. 
The topology of DNA in the chromosome has been further studied in models taking into account possible local modifications and global constraints \cite{barbi2005,barbi2010}. For reviews on transcription and transcriptional regulation in relation to epigenetic phenomena, see \cite{wolffe2000,berger2007}. It has early been understood that the topology of DNA is essential to understanding transcription, in prokaryotes, one suggestion for this is the possible supercoiling of DNA during transcription \cite{gamper1982,liu1987}. 

The mechanics of DNA and its topological supercoiling has been described using elastic models \cite{benham2005,thompson2002}. An elastic model has also been used to model DNA loops which play a role for the mechanics of transcription~\cite{balaeff2006}; one suggestion is that transcription of a certain gene can be repressed or activated whenever a DNA loop is formed that contains that gene \cite{saiz2006}. 
In the following, we use an entirely geometrical approach and model DNA as a double helix of two flexible tubes with fixed thickness. Under the presented assumptions, our main result is that for transcription to be possible, the pitch angle of the double helix must be constrained to a certain range, based on a differential length argument. Constraints that involve the degree of twisting can have two origins. 
One origin is of topological nature, such as conservation of the linking number \cite{skjeltorp1996}. Another origin is longer-range interactions that are due to the range of the involved forces 
in biological helices \cite{french2010,kornyshev2007}.

An aspect of transcription involving the pitch angle is the various stresses related to twist and strain when the B-DNA reorganizes itself prior to transcription. 
Experimental studies using magnetic tweezers and optical trapping suggest that DNA has a negative strain-twist coupling \cite{lionnet2006,gore2006}.  Elongated DNA will therefore tend to rotate through a larger angle per set of base-pairs, i.e. to be winding-up.
Recently, an analysis of the observed stick-slip melting of DNA under tension has revealed a zig-zag-like dynamics 
\cite{gross2011}. In Ref.~\cite{olsen2010}, we have suggested a geometrical model that describes the phenomenon of winding-up of biological double helices as a generic phenomenon for all molecular double helices for which the pitch angle is below a specific value.

\begin{figure}[t]\centering
\includegraphics[width=6.1cm]{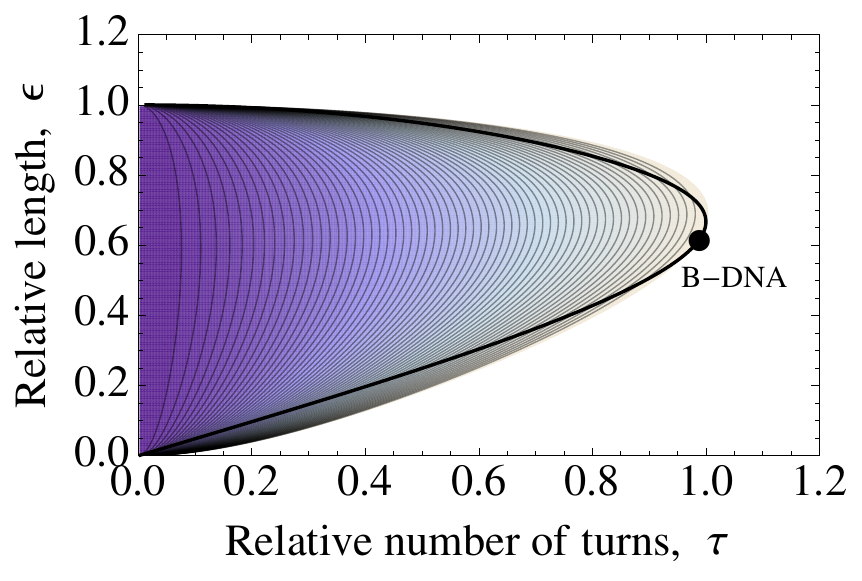}
\caption{\it Relative length, $\epsilon$,  of double helix as a function of the relative number of turns, $\tau$, on the strand. The curve is plotted for the asymmetric case, $\Psi=2.513$. The solid curve is the configurations where the two strands are in contact. The pitch angle $v_\bot$ increases along the curve from $0^\circ$ at the lower left corner, to $41.8^\circ$ at the turning point, and finally to $90^\circ$ at the upper left end-point. In the present paper B-DNA is found to have a pitch angle of $\sim 38^\circ$ corresponding to the point below the turning point. The contour lines have constant volume fraction, along the outer contour its value is 0.597 corresponding to B-DNA.} 
\label{fig:1}
\end{figure}

In the following, we review the geometry used in the analysis and study its implication on the pitch angle. Later, we compare with the experimentally available data and determine the pitch angle of B-DNA. 
Twisting two strands together to form e.g. a geometry akin to the double-stranded DNA, the resulting double helical structure has a length, $L_r(n)$, that depends on the number of twists, $n$ \cite{bohr2011a}. When the strands are parallel to each other, the length
of the double helix is the same as the length of the strands, and the double helix becomes shorter and shorter as turns are added. Only up to a maximum number of turns can be added, resulting ultimately in a geometry which is maximally rotated. In addition, this structure has a vanishing strain-twist coupling  \cite{olsen2010}. As rotations are removed from the helix, the length becomes further reduced. Therefore, $L_r(n)$ is shaped like a horizontal hairpin, see Fig.~\ref{fig:1}. 

In the detailed calculation of $L_r(n)$, the double helix is simplified to be consisting of two circular tubes of diameter $D$, whose center lines are coiled with pitch $H$ on a common cylinder of radius $a$. 
The pitch angle, $v_\bot$, measured from the horizontal is determined by $\tan v_\bot = H/2\pi a$. For these idealized structures, the steric interactions of molecular DNA are described by letting the tubes be in contact with hard walls, i.e. the distance between the two helical lines should satisfy 

\begin{equation}
\min (|\vec{r}_1(t_1)-\vec{r}_2(t_2)|) =D\, ,
\label{eq:min}
\end{equation}
%For $v_\bot=45^\circ$, and above, $W=2D$ \cite{pieranski1998}. 

\noindent where $\vec{r}_1, \vec{r}_2$ describe the two helical center lines. 

For transcription we consider DNA to be part of a structure that does not change on a large scale. 
Within this structure we consider an intermediate structure which is allowed to change while being maintained as a double helix. Finally, we consider a relatively short stretch of DNA that separates (melt) to single-stranded DNA. The first of these considerations require that the length of DNA during transcription, $L_{DNA}'$, to be greater than the length of DNA before transcription, $L_{DNA}$, i.e.

\begin{equation}
L'_{DNA} > L_{DNA}\, .
\label{eq:length}
\end{equation}
 
 \noindent Now, the twisting behavior of DNA depends on its pitch angle, $v_\bot$. If the pitch angle is such that it would just unwind under strain, then Eq. (\ref{eq:length}) would not be fulfilled, as shown below. Instead the DNA would tend to get shorter during transcription. This is the origin of the constraint on $v_\bot$ and shows that the behavior of B-DNA under strain must be opposite of that of a rope.

Mathematically, the two helical center lines in Eq. (\ref{eq:min}) have the parametric equations,

\begin{eqnarray}
\vec{r}_1(t_1) &=& (a\cos t_1, a \sin t_1, (H/2\pi)t_1)\, , \nonumber\\ 
\vec{r}_2(t_2) &=& (a\cos t_2, a \sin t_2, (H/2\pi)t_2+\Psi (H/2\pi))\, ,
\label{eq:0}
\end{eqnarray}

\noindent with $t_1, t_2 \in {\bf R}$. The parameter
$\Psi$ is a phase parameter that determines if the double helix is symmetric or asymmetric.
For a symmetric double helix, e.g. A-DNA, $\Psi$ is equal to $\pi$ and for an asymmetric double helix, akin to B-DNA, we have  $\Psi =\pi 144^\circ / \, 180^\circ  = 2.513$. The phase difference of 144$^\circ$ for the minor groove in B-DNA is described in Ref. \cite{klug1981}. With $D$ being the tube diameter, the points of tube-tube contacts obey the equation

\begin{equation}
D^2 =  a^2(\cos t -1)^2+a^2 \sin^2 t + \left( \frac{H}{2\pi}\right)^2(\Psi +t)^2\, ,
\label{eq:1}
\end{equation} 

\noindent where $t = t_1-t_2$.
The solutions to the condition of minimum distance, Eq. (\ref{eq:min}), is found from a numerical study of the equation, $d D^2/dt=0$; for further details, see Refs.~\cite{olsen2009, przybyl2001,neukirch2002}. These solutions have recently been expressed in a non-Euclidian geometry \cite{bruss2012}.

Figure~\ref{fig:1} shows how the (relative)  length of the double helix depends on the number of turns in the twisted strands using Eq. (\ref{eq:1}); the specific curve is for the asymmetric double helix with $\Psi=2.513$.
The vertical axis is the relative length, $\epsilon$, of the double helix, where
$\epsilon =L_r/L_s$ and $L_s$ is the strand length. The horizontal axis is the relative number of turns, $\tau=n/n_{max}$, where $n_{max}$ is the number of turns at the turning point for $L_s=1$. At the turning point, where $\tau$ is maximal, the pitch angle of the double helix takes a unique value, $v_\bot=v_{ZT}$, at which the double helix has zero strain-twist coupling. In \cite{olsen2010} it was found that 
$v_{ZT}=41.8^\circ$ for an asymmetric double helix (B-DNA) corresponding to $\epsilon=0.6665$. 
The contour plot in Fig. \ref{fig:1} shows lines of constant volume fraction, $f_V$. The volume fraction is defined locally as the volume of the two strands relative to the volume of a smallest enclosing cylinder \cite{olsen2009}. The boundary of the contour plot has $f_V=0.597$ which is the maximal volume fraction for the asymmetric double helix. The contour plot shows that when straining DNA, the geometry will tend to follow the solid curve as its tangent also is tangent to lines of constant volume fraction.

A double helix structure with a pitch angle greater than the zero-twist angle, $v_\bot\geq v_{ZT}$, is at the upper part of the curve in Fig.~\ref{fig:1}.
Now, there are two possibilities for the DNA double helix. Either $v_\bot \geq v_{ZT}$, or $v_\bot < v_{ZT}$. We will argue that actually,  $v_\bot < v_{ZT}$, so that DNA cannot be a maximally twisted structure, and that this bound is in agreement with the principle of close-packing of helices \cite{olsen2009}.
Let us assume that transcription is to take place on a short stretch of a long stretch of DNA without changing the total twist. 
The long DNA has strand length $L$, and the short stretch strand length $l$. The number of turns on the long DNA is
$n_{DNA} = \tau_L \cdot n_{max}\cdot L$, and during transcription it is
$n'_{DNA} =  \tau_{L-l} \cdot n_{max}\cdot (L-l)$.
Therefore, $\tau_{L-l} = \frac{L}{L-l} \tau_L$. On the stretch where transcription takes place, we have $\epsilon_l = 1$; that corresponds to the upper left endpoint of the curve in Fig.~\ref{fig:1}. The remaining part of the DNA has a relative length $\epsilon_{L-l}$, which for now is undetermined. The difference in the length of DNA after and before transcription is,

\begin{equation}
L_{DNA}' -L_{DNA}= l\cdot \epsilon_l + (L-l)\cdot \epsilon_{L-l} - L\cdot \epsilon_L\ .
%= L(\epsilon_{L-l}-\epsilon_L) + l(1-\epsilon_{L-l})\, .
\label{eq:2}
\end{equation} 

\noindent If this number is positive, then the DNA gets longer under transcription. If it is negative the DNA gets shorter, which is not generally physically possible. We assume that $l$ is small compared to $L$, $l \ll L$, and series expand $\epsilon_{L-l}$ to obtain, $\epsilon_{L-l}= \epsilon_l + 
\frac{\partial \epsilon}{\partial \tau}\frac{L}{L-l}\tau_L$. 
Therefore Eq. (\ref{eq:2}) can be rewritten as

\begin{equation}
L_{DNA}' -L_{DNA}= L\left( 1-\epsilon_L + \frac{\partial \epsilon}{\partial \tau}\tau_L
\right)\, .
\end{equation} 

\noindent Due to the convex nature of the curve in Fig.~\ref{fig:1}, the requirement that this difference is greater than zero cannot be obtained on the curves upper part where it is continuously bending downwards. We must be on the lower part of the curve in Fig.~\ref{fig:1}, so that the pitch angle obeys $v_\bot < v_{ZT}$. 

What is the experimentally determined pitch angle, $v_\bot$, of B-DNA? 
There is some debate in the scientific community about the interpretation and specific value of the pitch angle of B-DNA. 
In particular, some publications advocate a structure for DNA that has a pitch angle of 45$^\circ$ \cite{stasiak2000,poletto2008} based on the {\it symmetric} double helix. 
This corresponds to a point on the upper part of the A-DNA symmetric version of the curve in Fig.~\ref{fig:1} with $\epsilon = 0.707$. We will argue that the pitch angle for B-DNA is about 38$^\circ$.
We determine the pitch angle of DNA in the following way. The DNA pitch height, $H$, can accurately be determined by experiment. It is the height for which the helical structure progresses with precisely one full rotation. The helix-packing diameter \cite{dickerson1982}, or simply the width, $W$, of the molecule, can also be accurately determined. From $H$ and $W$ the aspect ratio can be calculated as, 

\begin{equation}
A=H/W\, .
\end{equation}

\noindent A classical work that lists the values of $H$ and $W$ for crystal structures of 
DNA is the Cold Harbor Symposium contribution by Dickerson {\it et al.}
\cite{dickerson1982}. The results are reproduced in Table~\ref{tab:1}, together with our calculated aspect ratios and the obtained pitch angles.
\begin{figure}[t]\centering
\includegraphics[width=4.1cm]{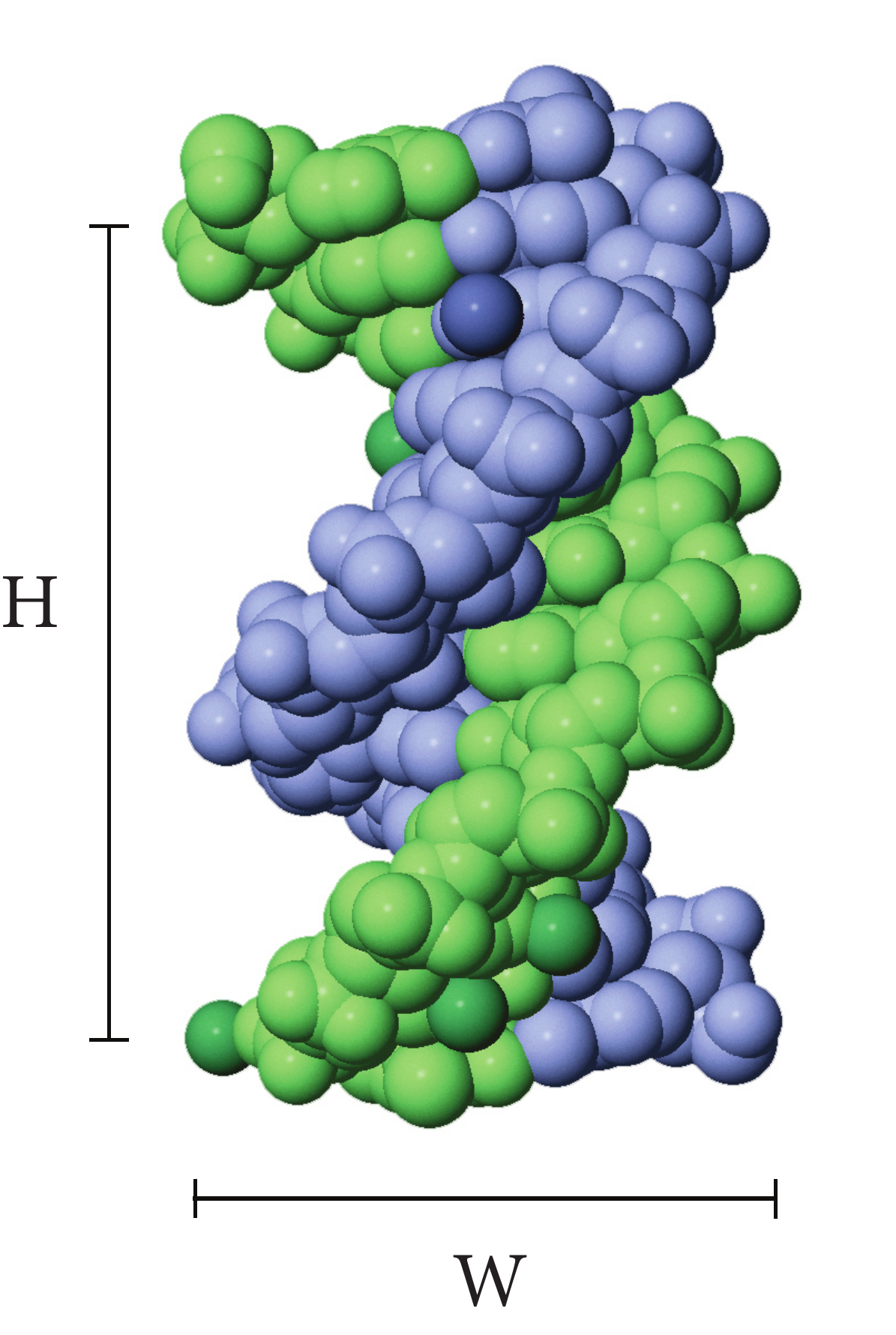}
\caption{\it B-DNA structure PDB $425D$ \cite{pdb425D}.} 
\label{fig:2}
\end{figure}
Figure~\ref{fig:2} depicts $H$ and $W$ for the B-DNA structure PDB 425D \cite{pdb425D}. 
The helical geometry is clearly visible, however it is superficial to determine the pitch angle by the naked eye from a two-dimensional projection. One needs to carry out a genuine three-dimensional geometrical study.

\begin{figure}[t]\centering
\includegraphics[width=6.1cm]{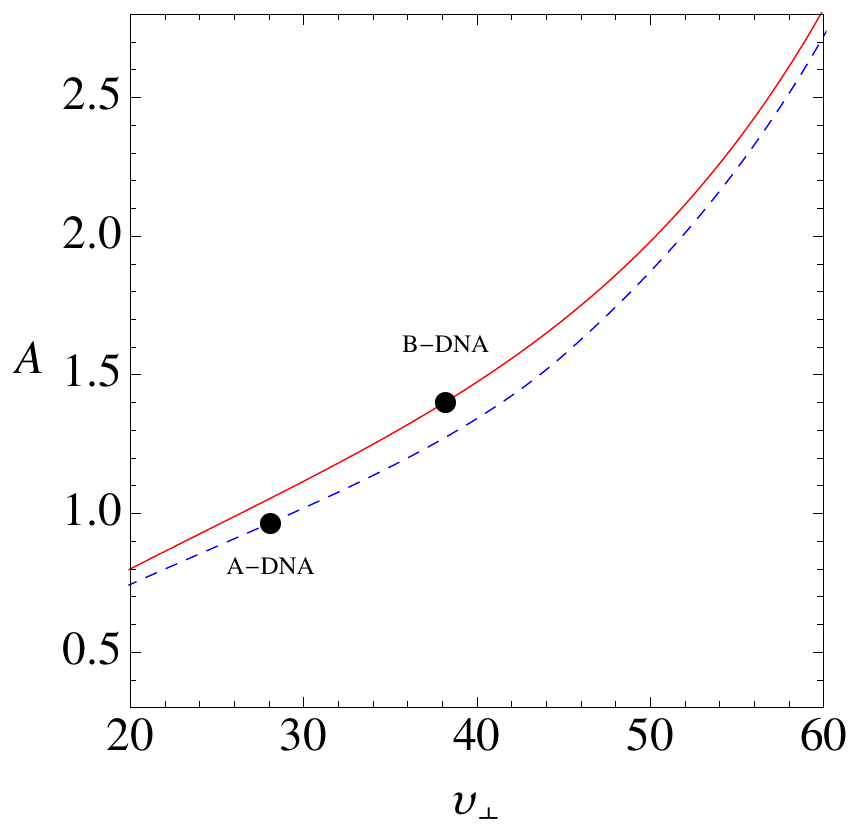}
\caption{\it Aspect ratio $A$ versus pitch angle $v_\bot$ plotted for double helix with minor groove (solid, red curve) and symmetric double helix (dashed, blue curve). The solid circles are for A-DNA, and B-DNA using Ref. \cite{dickerson1982}.}
\label{fig:3}
\end{figure}

The pitch angle can be indirectly determined from the aspect ratio above. 
For the double helix with phase parameter $\Psi$, the equation (\ref{eq:1}) gives a unique curve for $2a/D$ as a function of pitch angle, that is monotonically decreasing. The relation between the pitch angle $v_\bot$ and the aspect ratio $A$ reads,

\begin{equation}
\label{aspect}
A(v_\bot)= \frac{\pi \tan v_\bot}{\phi (v_\bot)}\, ,
\end{equation} 

\noindent where $\phi (v_\bot) = 1+ D/2a$. Figure~\ref{fig:3} shows how the aspect ratio, $A$, depends on the pitch angle, $v_\bot$, for a tubular double helix by solving numerically Eq. (\ref{eq:min}). 
The solid curve in Fig.~\ref{fig:3} is for an asymmetric double helix mimicking B-DNA. For an aspect ratio of 1.40 we obtain the pitch angle $v_\bot = 38^\circ$ shown by a solid circle. 
The dashed curve is for a symmetric double helix such as A-DNA. It passes through the point $(45^\circ,\pi/2)$. 
The experimentally determined aspect ratio for A-DNA is $0.96$. This corresponds to a pitch angle of 28$^\circ$ as shown in Fig.~\ref{fig:3} by a second solid circle. 

In its crystallized form, B-DNA has 10 base-pairs per turn, and using Eq. (\ref{aspect}) it gives a result for the pitch angle which is $\sim38^\circ$. Non-crystallized DNA, on the other hand, has about 10.5 base-pairs per turn. It is therefore further away from the turning point of the curve in Fig. \ref{fig:1} and consequently has a slightly lower pitch angle \cite{notepitchangle}. 
Table~1 summarizes our results for crystallized DNA.
\begin{table}[h!]\center
\begin{tabular}{llllll} 
\hline\noalign{\smallskip} 
&  $H$[\AA] &  $W$[\AA]  &  $A$ &  $v_\bot$ & $v_{ZT}$\\  [0.5ex]
\noalign{\smallskip}\hline\noalign{\smallskip}
B-DNA & 33.2  & 23.7  & 1.40 & 38$^\circ$ & 41.8$^\circ$\\  [1.0ex]
A-DNA & 24.6  & 25.5  & 0.96 & 28$^\circ$ & 39.4$^\circ$\\ [0.5ex]
\noalign{\smallskip}\hline
\end{tabular}
\caption{\it Pitch height, $H$, and width, $W$, experimental values for crystallized A-DNA and B-DNA reproduced from Ref. \cite{dickerson1982}. Also given are derived values for aspect ratio $A$ and pitch angle $v_\bot$, and the corresponding zero-twist pitch angle $v_{ZT}$ from Ref. \cite{olsen2010}.}  
\label{tab:1}
\end{table}
%Numerical values are given in {\bf Supplementary Table~1}, which lists aspect ratios for symmetric and asymmetric double helices in the range $v_\bot = 25^\circ$ to $45^\circ$.

\noindent The obtained pitch angle for DNA is therefore in agreement with the bound determined from transcription.
It can however seem a surprising result, how can the DNA pitch angle be below 45$^\circ$? In the symmetric double helix of $\Psi=\pi$, the two strands have a straight line of contact for $v_\bot\geq45^\circ$ \cite{pieranski1998}. Outside this interval, the two strands have a helical line of contact and the double helix has a central channel.
In Ref.~\cite{neukirch2002}, an elastic model is discussed where the pitch angle is to be greater than or equal to 45$^\circ$, which is suggested to be a lock-up angle for a two-ply \cite{thompson2002}. For the purpose of predicting the molecular interactions underlying the value of the pitch angle of DNA an elastic model with energy terms for curvature and torsion is not in agreement with the proposed bound. Presumably such an elastic model does not describe well the interactions at play on the molecular level.

Both of the empirical numbers for $v_\bot$ are in good agreement with the criterion suggested in \cite{olsen2009} that the dominating packing feature of DNA is determined by the obtainable volume fraction. I.e. that the criterion is consistent with an optimum use of available space. The packing of helical DNA therefore follows a classic packing rational in condensed matter \cite{krishna1981}. It seems  pertinent to pursue optimal packing in terms of maximization of the volume fraction, and hence density. 
The theoretical optimal values using the volume-packing criterion is 32.5$^\circ$ for the symmetric helix 
\cite{olsen2009,ohara2011} and 38.3$^\circ$ for the asymmetric helix \cite{olsen2009}, both less than the corresponding zero-twist angles so that transcription can take place.

\subsection*{Acknowledgements}
This work is supported by the Villum Foundation.


\begin{thebibliography}{}

\bibitem{becavin2010}
C. Becavin, M. Barbi, J-M Victor, and A. Lesne,
Transcription within Condensed Chromatin: Steric Hindrance Facilitates Elongation,
{\it Biophysical Journal} {\bf 98}, 824-833 (2010).

\bibitem{barbi2005} 
M. Barbi, J. Mozziconacci, J.M. Victor,
How the chromatin fiber deals with topological constraints,
{\it Phys. Rev. E.}, {\bf 71}, 031910 (2005).

\bibitem{barbi2010}
M. Barbi, J. Mozziconacci, J.M. Victor, H. Wong, C. Lavelle,
On the topology of chromatin fibres,
{\it Interface Focus} {\bf 2}, 546-554 (2012).

\bibitem{wolffe2000}
A.P. Wolffe, D. Guschin,
Review: chromatin structural features and targets that regulate transcription,
{\it Journal of structural biology} {\bf 129},102-122 (2000).

\bibitem{berger2007}
S.L. Berger, 
The complex language of chromatin regulation during transcription,
{\it Nature} {\bf 447}, 407Ð412 (2007).

\bibitem{gamper1982}
H.B. Gamper, J.E. Hearst, 
A topological model for transcription based on unwinding angle analysis of E. coli 
RNA polymerase binary, initiation and ternary complexes,
{\it Cell} {\bf 29}, 81-90 (1982). 

\bibitem{liu1987}
L.F. Liu, J.C. Wang, 
Supercoiling of the DNA template during transcription,
{\it Proc. Natl. Acad. Sci USA} {\bf 84}, 7024-7027 (1987).

\bibitem{benham2005}
C.J. Benham and S.P. Mielke,
DNA mechanics,
{\it Annual Review of Biomedical Engineering}
{\bf 7}, 21-53 (2005).

\bibitem{thompson2002}
J.M.T. Thompson, G.H.M. van der Heijden, and S. Neukirch,
Supercoiling of DNA plasmids: mechanics of the generalized ply,
{\it Proc. R. Soc. Lond.} A {\bf 458}, 959-985 (2002).

\bibitem{balaeff2006}
A. Balaeff, L. Mahadevan, and K. Schulten, 
Modeling DNA loops using the theory of elasticity,
{\it Phys. Rev. E} {\bf 73}, 031919 (2006).

\bibitem{saiz2006}
L. Saiz, J.M.G. Vilar,
DNA looping: the consequences and its control,
{\it Curr. Opin. Struct. Biol.} {\bf 16}, 344Ð350 (2006).

\bibitem{skjeltorp1996}
A.S. Skjeltorp, S. Clausen, G. Helgesen, P. Pieranski, "Knots and applications to biology, 
chemistry and physics", in: Physics of biomaterials: Fluctuations, Selfassembly, and Evolution, eds Riste T,
Sherrington DC NATO ASI Series E 322 pp 187-217. (Kluwer Academic Publishers, Dordrecht, 1996)

\bibitem{french2010}
R.H. French {\it et al.},
Long range interactions in nanoscale science,
{\it Rev. Mod. Phys.} {\bf 82}, 1887-1944 (2010). 

\bibitem{kornyshev2007}
A.A. Kornyshev, D.J. Lee, S. Leikin, and A. Wynveen,
Structure and interactions of biological helices
{\it Rev. Mod. Phys.} {\bf 79}, 943 (2007).

\bibitem{lionnet2006}
T. Lionnet, S. Joubaud, R. Lavery, D. Bensimon, V. Croquette,
Wringing out DNA,
{\it Phys. Rev. Lett.} {\bf 96}, 178102 (2006).

\bibitem{gore2006}
J. Gore, M. Bryant, M. N\"{o}llmann, M. U. Le, N. R. Cozzarelli, C. Bustamante,
DNA overwinds when stretched,
{\it Nature} {\bf 442}, 836 (2006).

\bibitem{gross2011}
P. Gross, N. Laurens, L. Oddershede, U. Bockelmann, E.J.G. Peterman, G.J.L. Wuite, 
Quantifying how DNA stretches, melts and changes twist under tension, 
{\it Nature Physics} {\bf 7}, 731-736 (2011).

\bibitem{olsen2010}
K. Olsen and J. Bohr,
The geometrical origin of the strain-twist coupling in double helices, 
{\it AIP Advances} {\bf 1} 012108 (2011).

\bibitem{bohr2011a}
J. Bohr and K. Olsen, 
The ancient art of laying rope,
{\it EPL} {\bf 93} 60004 (2011).

\bibitem{klug1981}
A. Klug, L.C. Lutter,
The helical periodicity of DNA on the nucleosome,
{\it Nucleic Acids Res.} {\bf 9}, 4267 (1981).

\bibitem{olsen2009}
K. Olsen, and J. Bohr,
The generic geometry of helices and their close-packed structures,
{\it Theor. Chem. Acc.} {\bf 125}, 207-215 (2010).

\bibitem{przybyl2001}
S. Przyby\l ,  and P. Piera\'{n}ski,
Helical close packings of ideal ropes,
{\it Eur. Phys. J. E} {\bf 4}, 445-449 (2001).

\bibitem{neukirch2002}
S. Neukirch, and G.H.M. van der Heijden,
Geometry and mechanics of uniform $n$-plies: from engineering ropes to biological filaments,
{\it Journal of Elasticity} {\bf 69}, 41-72 (2002).

\bibitem{bruss2012}
I.R. Bruss, G.M. Grason,
Non-Euclidean geometry of twisted filament bundle packing,
{\it PNAS} {\bf 109}, 10781-10786 (2012).

\bibitem{stasiak2000}
A. Stasiak, J.H.  Maddocks,
Mathematics: Best packing in proteins and DNA,
{\it Nature} {\bf 406}, 251-253 (2000).

\bibitem{poletto2008}
C. Poletto, A. Giacometti, A. Trovato, J. R. Banavar, and A. Maritan,
Emergence of secondary motifs in tubelike polymers in a solvent, 
{\it Phys. Rev. E} {\bf 77}, 061804 (2008).

\bibitem{dickerson1982}
R.E. Dickerson, H.R. Drew, B.N. Conner, M.L. Kopka, and P.E. Pjura,
Helix Geometry and Hydration in A-DNA, B-DNA, and Z-DNA,
{\it Cold Spring Habor Symp. Quant. Biol.} {\bf 47}, 13-24 (1983).

\bibitem{pdb425D}
H. Rozenberg, D. Rabinovich, F. Frolow, R.S. Hegde, Z. Shakked,
Structural code for DNA recognition revealed in crystal structures of papillomavirus E2-DNA targets,
{\it Proc. Natl. Acad. Sci. USA} {\bf 95}, 15194 (1998).

\bibitem{notepitchangle}
For 10.5 bp$/2\pi$ in B-DNA, the number of turns, $\tau$, is reduced by a factor of $10/10.5$, hence a quick estimate finds that the pitch angle is $v_\bot \sim 33^\circ$, i.e. to be further below the zero-twist angle. Other sources, such as textbooks, differ slightly on the reported structural parameters and a typical number given is $2a= 20$~\AA. This corresponds to a pitch angle of $v_\bot\sim36^\circ$.

\bibitem{pieranski1998}
P. Piera\'{n}ski and S. Przyby\l,
{\it In search of ideal knots},
in "Ideal Knots", pp 20--41,
Editors A. Stasiak, V. Katritch, and L.H. Kauffman,
(World Scientific, Singapore, 1998). 

\bibitem{krishna1981}
P. Krishna and P. Pandey,
{\it Close-Packed Structures},
Wales: International Union of Crystallography,
(University College Cardiff Press, 1981).

\bibitem{ohara2011} J. O'Hara , 
Ideal, best packing, and energy minimizing double helices,
{\it Prog. Theor. Phys. Supplement} {\bf 191}, 215-224 (2011).

\end{thebibliography}
\end{document}